%% file: main.tex
\documentclass[letterpaper, 10 pt, conference]{ieeeconf}  % Comment this line out
\IEEEoverridecommandlockouts                              % This command is only
\let\labelindent\relax
\usepackage[utf8]{inputenc}
\DeclareUnicodeCharacter{221E}{$\infty$}
\usepackage{graphicx} % for pdf, bitmapped graphics files
\usepackage{amsmath} % assumes amsmath package installed
\usepackage{amssymb}  % assumes amsmath package installed
\usepackage{float}

\usepackage{amsfonts}
\usepackage{amsthm}
\usepackage{mathtools}
\usepackage{bm}
\usepackage{xcolor}
\usepackage{cite}
\usepackage{hyperref}
\usepackage{enumitem}
\usepackage{booktabs}
\usepackage{multirow}
\usepackage{array}
\usepackage{algorithm}
\usepackage{algpseudocode}
\usepackage{microtype}
\usepackage{placeins}

\newtheorem{definition}{Definition}
\newtheorem{theorem}{Theorem}
\newtheorem{lemma}{Lemma}
\newtheorem{proposition}{Proposition}

\theoremstyle{remark}
\newtheorem{remark}{Remark}

\title{\LARGE \bf
Verifiably Safe Decentralized Control\\ via Event-Triggered Topology Switching}

\author{Amber He Wei, Shiyu Cheng, Andrew Clark, and Bruno Sinopoli
\thanks{A. H. Wei, S. Cheng and A. Clark are with the Department of Electrical and Systems Engineering, Washington University in St. Louis, St. Louis, MO, USA. Email: \{hwei, cheng.shiyu, andrewclark\}@wustl.edu}
\thanks{B. Sinopoli is with the School of Electrical, Computer and Energy Engineering, Arizona State University, Tempe, AZ, USA. Email: bruno.sinopoli@asu.edu}
}

\begin{document}

\maketitle
\thispagestyle{empty}
\pagestyle{empty}

%%%%%%%%%%%%%%%%%%%%%%%%%%%%%%%%%%%%%%%%%%%%%%%%%%%%%%%%%%%%%%%%%%%%%%%%%%%%%%%%
\begin{abstract}
This paper proposes a decentralized event-triggered topology-switching protocol for multi-agent systems under estimation uncertainty. We consider an intermittent communication setting in which each agent switches between connected and disconnected modes and makes communication decisions using locally computable certificates, while the closed-loop state is required to remain safe. The proposed protocol follows a safety-driven triggering principle in which agents reconnect before their certificates reach the safety boundary and disconnect only after recovering to a stricter bound below the safety limit. Our design enables intermittent connectivity while maintaining the estimation error within a certified safety bound and excluding Zeno behavior. We evaluate the proposed protocol on a quadruple-tank example, showing that it keeps all tank deviations within the safety bound while allowing agents to remain disconnected for most of the run.
\end{abstract}

%%%%%%%%%%%%%%%%%%%%%%%%%%%%%%%%%%%%%%%%%%%%%%%%%%%%%%%%%%%%%%%%%%%%%%%%%%%%%%%%
\input{Intro}
\input{Model}
\input{Formulation}
\input{safety}
\input{Trigger}

\input{Simulation}
\input{Conclusion}

\bibliographystyle{IEEEtran}
\bibliography{references}
\end{document}

%% file: Intro.tex
\section{Introduction}
\label{sec:intro}
Multi-agent systems, such as drone swarms and autonomous ground vehicle platoons, are widely deployed in applications that require coordinated action across a wide area \cite{bullo2009distributed}. Distributed control laws have been proposed for multi-agent systems to satisfy desired properties such as formation maintenance, target tracking, stability, and safety. The control laws typically require periodic state information exchange between neighboring agents, and thus incur significant communication overhead.

Event-triggered communication and control has been proposed to mitigate communication overhead \cite{ge2021dynamic}. In an event-triggered framework, agents do not communicate at a fixed rate, but instead broadcast their local state information when certain triggering conditions are satisfied, e.g., when the agents are close to violating stability or safety constraints. This reduces utilization of the communication channel. Event-triggered control algorithms have been proposed for consensus \cite{cui2023decentralized,nowzari2019event}, formation stability \cite{an2023event,peng2013event}, tracking moving targets \cite{deng2022distributed}. Event triggered Control Barrier Functions (CBFs) have been proposed \cite{wang2024control,luo2020multi,sabouni2023optimal} to certify safety, defined as forward invariance of a given safe set. 

While these approaches show significant promise, they retain some sources of vulnerability and inefficiency. First, even when not transmitting, the agents must remain connected to an underlying communication network, which incurs overhead due to periodic neighbor discovery and topology control protocols \cite{rajaraman2002topology}. Second, even though agents transmit less frequently, they must still listen for and decode messages broadcasted by their neighbors, which consumes scarce energy resources. Finally, the static network topology remains vulnerable to malicious attacks such as denial-of-service.

Recently, the authors of \cite{10360262} have proposed a different framework for multi-agent coordination, namely, \emph{intermittent, event-triggered connectivity}. In this framework, agents connect to the network only when triggering conditions are satisfied, and exchange information at a fixed rate while connected. This approach was shown in~\cite{10360262} to guarantee robust asymptotic convergence to a desired ellipsoidal region. 

In this paper, we investigate the problem of guaranteeing safety of multi-agent systems with intermittent, event-triggered network topologies.  Different from \cite{10360262}, where state estimation errors may grow outside the target region while an agent is disconnected as long as the error recovers after reconnection, we require the certified error bound to hold for all time, including during disconnected intervals. This leads to a safety-driven triggering design in which agents reconnect before their local certificates reach the safety limit and disconnect only after recovering below a stricter bound.

In our approach, each agent maintains a Luenberger-type observer of the joint state space which can operate in either a \emph{connected mode} in which the agent is communicating with its neighbors or a \emph{disconnected mode} in which the agent updates its estimate using only locally-available information. The goal of the agents is to satisfy a safety constraint on the joint agent states that is formulated as a barrier certificate. We make the following contributions:
\begin{itemize}
\item We propose a triggering protocol in which agents connect
independently using locally computable certificates, while connected agents
reachable over active links disconnect synchronously only when they
simultaneously satisfy an inner recovery condition.
\item We derive constraints on the triggering parameters that guarantee safety by ensuring both recovery under full connectivity and reconnection before safety violation.
\item We demonstrate the framework on a water-tank example, showing that local triggering certificates regulate connected and disconnected intervals while maintaining safety.
\end{itemize}

The paper is organized as follows. Section~\ref{sec:model} presents the system model and decentralized state-estimation framework. Section~\ref{sec:formulation} formulates the problem. Section~\ref{sec:safety} establishes the protocol design and safety guarantees. Section~\ref{sec:trigger} constructs the local triggering certificate. Section~\ref{sec:simulation} presents the simulation, and Section~\ref{sec:conclusion} concludes the paper.

%% file: Model.tex
\section{Model and Preliminaries}
\label{sec:model}

\subsection{Notation}
For any matrix $M$, $M^\top$ denotes its transpose. For symmetric matrices, $M\succ 0$ ($M\succeq 0$) indicates positive definiteness (positive semidefiniteness). The standard Euclidean norm is denoted by $\| \cdot \|$. For a set of vectors $z_1,\dots,z_N$, $\mathrm{col}(z_1,\dots,z_N)\triangleq [z_1^\top, \ldots, z_N^\top]^\top$ represents their stacked column vector.

\subsection{System Model}
In this work, we consider a system of $N$ agents indexed by $\mathcal{N} \triangleq \{1,\dots,N\}$.  The system follows a continuous-time linear time-invariant (LTI) dynamical model:
\begin{equation}\label{system}
\begin{aligned}
\dot{x}(t) &= Ax(t) + Bu(t) + w(t), \\
y(t) &= Cx(t) + v(t),
\end{aligned}
\end{equation}
where $x(t)\in\mathbb{R}^n$ is the global state, $u(t)\in\mathbb{R}^m$ is the input, $y(t)\in\mathbb{R}^p$ is the measurement output, $w(t)\in\mathbb{R}^{n}$ is the process disturbance, and $v(t)\in\mathbb{R}^{p}$ is the measurement noise. The system matrices $A\in\mathbb{R}^{n\times n}$, $B\in\mathbb{R}^{n\times m}$, and $C\in\mathbb{R}^{p\times n}$ are known and constant.
The global state and input are partitioned among the $N$ agents as
$x(t) = \mathrm{col}\bigl(x_1(t),\dots,x_N(t)\bigr), 
u(t) = \mathrm{col}\bigl(u_1(t),\dots,u_N(t)\bigr),$ where $x_i(t) \in \mathbb{R}^{n_i}$ with $\sum_{i=1}^N n_i = n$, and $u_i(t) \in \mathbb{R}^{m_i}$ with $\sum_{i=1}^N m_i = m$.
Each agent $i\in\mathcal{N}$ accesses a local measurement
$y_i(t) = C_i x(t) + v_i(t) \in \mathbb{R}^{p_i},$ where $C_i \in \mathbb{R}^{p_i \times n}$ and $v_i(t) \in \mathbb{R}^{p_i}$, such that $\sum_{i=1}^N p_i = p$. Stacking these yields the global vectors and matrices $y(t) = \mathrm{col}\bigl(y_1(t),\dots,y_N(t)\bigr)$, $C = \mathrm{col}\bigl(C_1,\dots,C_N\bigr)$, and $v(t) = \mathrm{col}\bigl(v_1(t),\dots,v_N(t)\bigr)$.
The disturbances $w(t)$ and $v(t)$ are assumed bounded within known compact, quadratically defined sets for all $t \ge 0$:
\begin{equation*}
\label{disturbance}
\begin{aligned}
\mathcal{W} &\triangleq \{w\in\mathbb{R}^n: w^\top Q_w w \le 1\}, \quad  Q_w \in \mathbb{R}^{n \times n}, \ Q_w \succ 0, \\
\mathcal{V} &\triangleq \{v\in\mathbb{R}^p: v^\top R_v v \le 1\}, \quad  R_v \in \mathbb{R}^{p \times p}, \ R_v \succ 0.
\end{aligned}
\end{equation*}

\subsection{Communication Model}
\label{subsec:communication-model}
The communication network among the $N$ agents is modeled by a time-varying, undirected graph $\mathcal{G}(t) = (\mathcal{N}, \mathcal{E}(t))$, where the node set $\mathcal{N}$ corresponds to the agents and the edge set $\mathcal{E}(t) \subseteq \mathcal{N} \times \mathcal{N}$ represents the active communication edges at time $t \ge 0$. We assume the existence of an underlying static, undirected, and connected graph $\overline{\mathcal{G}} = (\mathcal{N}, \overline{\mathcal{E}})$ that defines the maximum allowable communication topology, where $\overline{\mathcal{E}} \subseteq \mathcal{N} \times \mathcal{N}$ represents the set of all possible communication edges between the agents.

The communication model adopted in this work follows the intermittent, event-triggered connectivity paradigm: rather than maintaining a fixed network, each agent connects to the network only when a locally evaluated triggering condition requires it. This avoids the communication and resource overhead of remaining connected at all times.

Each agent can switch between two modes:
\begin{itemize}
    \item \textbf{Disconnected Mode:} Agent $i$ operates independently,
    updating its state estimate using only its local measurements.
    \item \textbf{Connected Mode:} Agent $i$ exchanges information with its
    active neighbors, i.e., the neighboring agents that are
    also in connected mode, and uses the received information to update its
    estimate.
\end{itemize}
To switch between the two modes, we design the triggering protocol in Section~\ref{sec:trigger}, where each agent monitors a locally computed upper bound on its estimation error. An agent connects to the network on its own when its bound reaches the safety limit, beyond which safety can no longer be guaranteed. During the connected mode, each agent signals that it is ready to disconnect once its bound decreases to a designed inner safety threshold, which provides a margin strictly below the safety limit. An agent returns to the disconnected mode only when it and all agents
reachable from it over active links simultaneously satisfy the ready-to-disconnect
condition; at that time, those agents disconnect from the network together.

For each agent $i \in \mathcal{N}$, its \emph{potential neighboring set} is the
neighborhood induced by the underlying graph $\bar{\mathcal{G}}$,
\begin{equation}\label{eq:potential-neighbors}
    \mathcal{N}_i \triangleq \bigl\{\, j \in \mathcal{N}\setminus\{i\} :
    (i,j) \in \bar{\mathcal{E}} \,\bigr\},
\end{equation}
i.e., the set of agents with which agent $i$ may communicate when it is connected to the network. The agents that agent $i$ can actually communicate with at a given time are those that are also connected to the network. Let $\Psi(t) \subseteq \mathcal{N}$ denote the set of agents connected to the network at time $t$. The \emph{active neighboring set} of agent $i$ at time $t$ then collects the potential neighbors of agent $i$ that are simultaneously connected:
\begin{equation}
    \mathcal{N}_i(t) \triangleq
    \begin{cases}
        \mathcal{N}_i \cap \Psi(t), & i \in \Psi(t), \\
        \emptyset, & i \notin \Psi(t),
    \end{cases}
\end{equation}
so that $\mathcal{N}_i(t) \subseteq \mathcal{N}_i$ for all $t \ge 0$.

Let $\mathcal{A}(t)=[a_{ij}(t)]\in\mathbb{R}^{N\times N}$ be the symmetric adjacency matrix of $\mathcal{G}(t)$, where \(a_{ij}(t)=1\) if edge $(i,j)$ is active and \(a_{ij}(t)=0\) otherwise, with $a_{ii}(t)=0$. Let $\mathcal{L}(t)=[\ell_{ij}(t)]\in \mathbb{R}^{N \times N}$ denote the associated graph Laplacian, where $\ell_{ii}(t)$ is the number of active neighbors of agent $i$.

The static Laplacian associated with the underlying graph $\overline{\mathcal{G}}$ is denoted by $\overline{\mathcal{L}} \in \mathbb{R}^{N \times N}$. Because $\overline{\mathcal{G}}$ is connected, $\overline{\mathcal{L}}$ is positive semi-definite with a simple eigenvalue at $0$ and a strictly positive algebraic connectivity $\lambda_2(\overline{\mathcal{L}}) > 0$. Since $\mathcal{G}(t)$ is a subgraph of $\overline{\mathcal{G}}$, it follows that $\mathcal{L}(t) \preceq \overline{\mathcal{L}}$ for all $t \geq 0$.

\subsection{Decentralized State Estimation}

Each agent applies a static state-feedback control law based on its local estimate $\hat{x}_i(t) \in \mathbb{R}^n$ of the global state: $u_i(t) = K_i \hat{x}_i(t),$ where $K_i\in\mathbb{R}^{m_i\times n}$ is a given local gain matrix. Stacking these yields the global nominal closed-loop matrix $A_{bk} \triangleq A+BK \in \mathbb{R}^{n \times n}$, where $K \triangleq [K_1^T, \ldots, K_N^T]^T \in \mathbb{R}^{m \times n}$. Let $L_i \in \mathbb{R}^{n\times p_i}$ denote the $i$-th block of the networked observer gain, and define
$L \triangleq [L_1~\cdots~L_N]\in\mathbb{R}^{n\times p}$. The gain $L$ is designed such that
$A-LC$ is Hurwitz.
Then, we define the
mode-dependent gain \(\bar{L}_i(\ell_{ii}(t))\in\mathbb{R}^{n\times p_i}\) used by agent $i$:
\[
\bar{L}_i(\ell_{ii}(t))
=
\begin{cases}
\hat L_i, & \ell_{ii}(t)=0,\\
N L_i, & \ell_{ii}(t)>0,
\end{cases}
\]
where the matrix $\hat{L}_i \in \mathbb{R}^{n \times p_i}$ is agent $i$'s local observer gain, constructed via an observability decomposition. Let $T_i \in \mathbb{R}^{n \times n}$ be a similarity transformation matrix such that:
\begin{equation}
\begin{aligned}
T_i^{-1} A T_i &= 
\begin{bmatrix} 
A_i^o & 0 \\ 
A_i^{21} & A_i^{\bar{o}} 
\end{bmatrix},~ 
C_i T_i = \begin{bmatrix} C_i^o & 0 \end{bmatrix},
\end{aligned}
\end{equation}
where the pair $(A_i^o, C_i^o)$ is observable. The local gain is then defined as $\hat{L}_i \triangleq T_i [L_i^o, 0]^\top$, where $L_i^o$ is designed such that $A_i^o - L_i^o C_i^o$ is Hurwitz. Then, the decentralized observer dynamics for agent $i$ are:
\begin{equation}
\begin{aligned}
\dot{\hat{x}}_i(t) ={}& A_{bk}\hat{x}_i(t) + \bar{L}_i(\ell_{ii}(t))\bigl(y_i(t) - C_i\hat{x}_i(t)\bigr) \\
& + \eta \sum_{j=1}^N a_{ij}(t)\bigl(\hat{x}_j(t) - \hat{x}_i(t)\bigr),
\end{aligned}
\end{equation}
where $\eta \ge 0$ is the network coupling gain, and $\ell_{ii}(t)$ is the $i$-th diagonal entry of $\mathcal{L}(t)$.

\subsection{Error Dynamics and Closed-Loop Dynamics}

We define agent $i$'s estimation error of the global state as $e_i(t) \triangleq x(t) - \hat{x}_i(t) \in \mathbb{R}^n$. The corresponding error dynamics are given by:
\begin{align*}
\dot{e}_i(t) ={}& \bigl(A_{bk} - \bar{L}_i(\ell_{ii}(t))C_i\bigr)e_i(t) + w(t) - \bar{L}_i(\ell_{ii}(t))v_i(t) \\
& - \sum_{j=1}^N \bigl(B_j K_j + \eta\,\ell_{ij}(t)\bigr)e_j(t).
\end{align*}
By stacking the agents' global-state estimation errors as $e(t) \triangleq \mathrm{col}(e_1(t), \dots, e_N(t)) \in \mathbb{R}^{Nn}$, we compactly express the stacked error dynamics as:
\begin{equation}
\begin{aligned}
\label{eq:dynamics-error}
\dot{e}(t) ={}& \underbrace{\bigl(F(\mathcal{L}(t)) - \eta(\mathcal{L}(t) \otimes I_n)\bigr)}_{A_e(\mathcal{L}(t))} e(t) \\
& + \mathcal{I} w(t) - J(\mathcal{L}(t)) v(t),
\end{aligned}
\end{equation}
where $\mathcal{I} \triangleq \operatorname{col}(I_n, \dots, I_n) \in \mathbb{R}^{Nn \times n}$,
\begin{equation*}
    J(\mathcal{L}(t)) \triangleq \operatorname{BlkDiag}\bigl(\bar L_1(\ell_{11}(t)), \dots, \bar L_N(\ell_{NN}(t))\bigr),
\end{equation*}
and $F(\mathcal{L}(t)) \in \mathbb{R}^{Nn \times Nn}$ is a block matrix with block entries $F_{ij}(\mathcal{L}(t)) \in \mathbb{R}^{n \times n}$ defined as:
\begin{equation*}
F_{ij}(\mathcal{L}(t)) \triangleq 
\scalebox{0.9}{$
\begin{cases}
A_{bk} - \bar{L}_i(\ell_{ii}(t))C_i - B_i K_i, & \text{if } i = j,\\[4pt]
-B_j K_j, & \text{if } i \neq j.
\end{cases}$}
\label{eq:Fij}
\end{equation*}
Finally, substituting the local control inputs $u_i(t) = K_i x(t) - K_i e_i(t)$ into the system model yields the global closed-loop state dynamics:
\begin{equation}\label{eq:closed_loop_dynamics}
\dot{x}(t) = A_{bk}x(t) - Ee(t) + w(t),
\end{equation}
where $E \triangleq \begin{bmatrix} B_1 K_1 & \cdots & B_N K_N \end{bmatrix} \in \mathbb{R}^{n \times Nn}$.

%% file: Formulation.tex
\section{Problem Formulation}
\label{sec:formulation}

In this paper, we guarantee safety for a multi-agent system under intermittent connectivity by enforcing forward invariance of a certified subset of the prescribed safe region. This is achieved by constraining the estimation error within a designed error-safe set, so that the system state remains safe in the presence of estimation uncertainty and disturbances. We further design an event-triggered communication protocol in which agents reconnect individually when their local certificates approach the error-safe boundary, and connected neighboring agents disconnect jointly after recovery. Communication is therefore activated only when necessary to preserve safety.

\subsection{Safety Requirement}
\label{subsec:safety-requirement}

The system is required to satisfy a safety constraint for all time, i.e. $x(t) \in \mathcal{C}$ for all $t \ge 0$, where $\mathcal{C}$ is a safe operating region. The set $\mathcal{C}$ is defined by a polynomial $h : \mathbb{R}^n \to \mathbb{R}$ as
\begin{equation}\label{eq:safe-region}
    \mathcal{C} \triangleq \{x \in \mathbb{R}^n : h(x) \ge 0\}.
\end{equation}
The initial state is assumed feasible, i.e., $x(0) \in \mathcal{C}$.

Certifying forward invariance of $\mathcal{C}$ is generally
intractable, since the safe region may be irregular, making the standard invariance condition difficult to impose. We therefore certify safety through a polynomial barrier function $b : \mathbb{R}^n \to \mathbb{R}$ that
defines a certifiable inner approximation of $\mathcal{C}$, namely the
zero-superlevel set
\begin{equation}\label{eq:barrier-set}
    \mathcal{B} \triangleq \{x \in \mathbb{R}^n : b(x) \ge 0\}
\end{equation}
satisfying the containment $\mathcal{B} \subseteq \mathcal{C}$ and including the initial state, $x(0) \in \mathcal{B}$. Since $\mathcal{B} \subseteq \mathcal{C}$, forward invariance of $\mathcal{B}$ ensures $x(t) \in \mathcal{C}$ for all $t \ge 0$. Forward invariance of $\mathcal{B}$ is certified using a control barrier function, which ensures safety through the forward invariance of a zero-superlevel set \cite{8796030}. For $\dot{x} = f(x)$ with
$f$ locally Lipschitz, let $\phi : \mathbb{R}^n \to \mathbb{R}$ be
continuously differentiable and $\mathcal{S} \triangleq \{x : \phi(x)
\ge 0\}$. If there is an extended class-$\mathcal{K}$ function $\alpha$
with
\begin{equation}\label{eq:cbf-cond}
    \nabla\phi(x)^\top f(x) \ge -\alpha\!\big(\phi(x)\big),
    \qquad \forall\, x \in \mathcal{S},
\end{equation}
then $\mathcal{S}$ is forward invariant, i.e. $x(0)\in\mathcal{S}$
implies $x(t)\in\mathcal{S}$ for all $t\ge0$ \cite{8796030}. Applying the barrier condition~\eqref{eq:cbf-cond} to $b$ along the
closed-loop dynamics~\eqref{eq:closed_loop_dynamics}, forward invariance
of $\mathcal{B}$ is guaranteed if
\begin{equation}\label{eq:cbf-closedloop}
    \nabla b(x)^\top \bigl(A_{bk}x - E e + w\bigr)
    + \alpha_b\bigl(b(x)\bigr) \ge 0
\end{equation}
holds for every $x \in \mathcal{B}$, where $\alpha_b$ is an extended
class-$\mathcal{K}$ function.

\subsection{Estimation-Error Certificate}
\label{subsec:error-certificate}

The CBF condition~\eqref{eq:cbf-closedloop}, used to guarantee forward invariance of $\mathcal{B}$, depends on the stacked estimation error $e$. Since the agents do not have access to the true state, $e$ is not exactly known and must therefore be confined to a bounded region. We now define the certificate that quantifies this region.

\begin{definition}
\label{def:structured_error_weighting_matrix}
Let $P_e = P_e^\top \succ 0$ be a weighting matrix for each agent's global-state estimation error, and define the global error weighting matrix $\bar P \triangleq I_N \otimes P_e.$ For the stacked estimation error
    \begin{equation}  e(t)=\operatorname{col}\big(e_1(t),\ldots,e_N(t)\big),\label{eq:stacked_global_state_estimation_error}
\end{equation}
where $e_i(t)=x(t)-\hat{x}_i(t)$, the matrix $\bar P$ assigns the same weight $P_e$ to every $e_i(t)$.
\end{definition}

\begin{definition}
\label{def:error_certificate}
For the stacked estimation error \eqref{eq:stacked_global_state_estimation_error}, define the quadratic certificate
\begin{equation}
    V_e(e(t)) := e(t)^\top \bar P e(t),
    \label{eq:error_certificate}
\end{equation}
and the error-safe set as its unit sublevel set,
\begin{equation}
    \Omega \triangleq \big\{ e\in\mathbb R^{Nn} : e^\top \bar P e \le 1 \big\}.
    \label{eq:error_safe_set}
\end{equation}
Maintaining $V_e(e(t)) \le 1$ is equivalent to $e(t) \in \Omega$.
\end{definition}

Since the global error $e(t)$ is not computable, each agent instead evaluates a locally computable certificate $U_i(t)$ that upper-bounds the global error certificate,
\begin{equation}\label{eq:Ui-bound}
    V_e(e(t)) \le U_i(t).
\end{equation}

By~\eqref{eq:Ui-bound}, $V_e(e(t)) = e(t)^\top \bar P e(t) \le U_i(t) \le 1$, so $e(t) \in \Omega$ by~\eqref{eq:error_safe_set}.

\subsection{Problem Statement}
\label{subsec:problem-statement}

With the safety requirement and the estimation-error certificate, the problem addressed in this work is to co-design the barrier function $b$ and the error weighting matrix $\bar P = I_N \otimes P_e$, together with the local certificate $U_i(t)$ and the triggering protocol they induce, such that:
\begin{enumerate}
    \item[(i)] \textbf{Safety under the protocol:} under the triggering protocol, the estimation error satisfies $e(t) \in \Omega$ for all $t \ge 0$; the certified set $\mathcal{B}$ is forward invariant; and consequently, since $\mathcal{B} \subseteq \mathcal{C}$, the closed-loop state satisfies $x(t) \in \mathcal{C}$ for all $t \ge 0$;
    \item[(ii)] \textbf{Local certificate:} there exists a locally computable certificate $U_i(t)$ satisfying the upper bound~\eqref{eq:Ui-bound}, so that each agent can evaluate its triggering condition using only local information;
    \item[(iii)] \textbf{Feasible weighting matrix:} There exists a feasible error-weighting matrix $\bar P$, as defined in Definition~\ref{def:structured_error_weighting_matrix}, that certifies the conditions in (i).
\end{enumerate}

%% file: safety.tex
\section{Safety Under an Arbitrary Local Error Certificate}
\label{sec:safety}

This section establishes the safety guarantees of the closed-loop system. We show that if the estimation error lies in the error-safe set $\Omega$, then the robust CBF condition~\eqref{eq:cbf-closedloop} holds, certified through a sum-of-squares (SOS) construction, and that under full connectivity the error-safe set $\Omega$ is forward invariant. At any time, the network is either partially or fully connected; establishing safety in each case therefore yields safety under the switching protocol for all $t \ge 0$. Throughout this section, the certificate $U_i(t)$ is treated as an arbitrary function satisfying only the upper bound~\eqref{eq:Ui-bound}; its explicit construction, computable from information available to agent $i$, is deferred to Section~\ref{sec:trigger}.

\begin{definition}[Triggering protocol with synchronized release]
\label{def:thresholds}
The protocol uses the safety threshold $1$ and an inner threshold
$\gamma_{\mathrm{off}}\in(0,1)$, which defines the inner error-safe set
\begin{equation}\label{eq:inner-set}
    \Omega_{\mathrm{off}}\triangleq\{e\in\mathbb{R}^{Nn}:V_e(e)\le\gamma_{\mathrm{off}}\}\subset\Omega.
\end{equation}
Each agent carries a readiness bit $\sigma_i\in\{0,1\}$,
\begin{equation}\label{eq:readiness}
    \sigma_i(t)=1
    \ \iff\
    i\in\Psi(t)\ \text{and}\ U_i(t)\le\gamma_{\mathrm{off}};
\end{equation}
that is, $\sigma_i(t)$ is set whenever agent $i$ is connected and its current
bound satisfies $U_i(t)\le\gamma_{\mathrm{off}}$, and is cleared as soon as
$U_i(t)>\gamma_{\mathrm{off}}$ again, so the bit is memoryless rather than
latched. Here $\mathcal{R}_i(t)\subseteq\Psi(t)$ denotes the set of connected
agents reachable from $i$ over active edges in $\mathcal{G}(t)$, i.e.\ those $j$
for which some path $i=v_0,v_1,\dots,v_\ell=j$ has every edge $(v_{r-1},v_r)$
active.
Mode transitions are
\begin{equation}\label{eq:trigger-rule}
    \begin{cases}
    \text{agent} ~i\ \text{connects}, & U_i(t)\ge 1,\\[3pt]
        \text{agent} ~i\ \text{disconnects}, & \sigma_j(t)=1\ \ \forall j\in\mathcal{R}_i(t).
    \end{cases}
\end{equation}
Connection is decided per agent; disconnection is applied simultaneously to all
agents in $R_i(t)$ and occurs only at an instant when every agent in $R_i(t)$
is ready to disconnect, i.e., when $U_j(t)\le \gamma_{\mathrm{off}}$ holds at
once for all $j\in R_i(t)$. Upon disconnection, $\sigma_j=0$ for all $j\in R_i(t)$.
\end{definition}

The protocol is implemented as follows. At each time $t$, every agent $i$ measures its output and computes the local error bound $U_i(t)$ from information available to it. If agent $i$ is disconnected and $U_i(t)$ reaches the safety level $1$, it connects to the network. If agent $i$ is connected, it sets its readiness bit to $\sigma_i=1$ whenever $U_i(t)\le\gamma_{\mathrm{off}}$ and resets it to $\sigma_i=0$ as soon as $U_i(t)>\gamma_{\mathrm{off}}$. Readiness bits can be included in the routing or topology-control messages over the active links, so that all agents in $\mathcal R_i(t)$ can evaluate the synchronized-release condition. All agents in $\mathcal R_i(t)$ disconnect simultaneously at the first instant
when all of these bits equal $1$, equivalently when
$U_j(t)\le\gamma_{\mathrm{off}}$ holds at the same time for every
$j\in\mathcal R_i(t)$. If any $U_j(t)$ rises above
$\gamma_{\mathrm{off}}$ before release, then the corresponding bit returns to
$0$, and all agents in $\mathcal R_i(t)$ remain connected. Finally, every agent updates its state estimate using the observer of its current mode and applies its local control input.

\subsection{Safety from the Error Certificate}

We first show that if the estimation error lies in the error-safe set $\Omega$, then the robust CBF condition~\eqref{eq:cbf-closedloop} holds. Let $\Sigma[\,\cdot\,]$ denote the cone of sum-of-squares (SOS) polynomials in the indicated variables; any $p \in \Sigma[\,\cdot\,]$ is globally nonnegative. Define $\Phi(e) \triangleq 1 - V_e(e)$, so that $e \in \Omega$ is equivalent to $\Phi(e) \ge 0$, and define
\begin{equation}\label{eq:gb}
    g_b(x,e,w) \triangleq \nabla b(x)^\top\bigl(A_{bk}x - Ee + w\bigr) + \alpha_b\bigl(b(x)\bigr).
\end{equation}

\begin{theorem}
\label{thm:sos-cbf}
Let $b(x)$ be polynomial. Suppose there exist SOS multipliers $s_b, s_\Phi, s_w, s_v \in \Sigma[x,e,w,v]$ such that
\begin{equation}\label{eq:sos-condition}
\begin{aligned}
    g_b(x,e,w) &- s_b\, b(x) - s_\Phi\, \Phi(e) - s_w\bigl(1 - w^\top Q_w w\bigr) \\&- s_v\bigl(1 - v^\top R_v v\bigr) \in \Sigma[x,e,w,v].
\end{aligned}
\end{equation}
Then the robust CBF condition~\eqref{eq:cbf-closedloop} holds for every $x \in \mathcal{B}$, $e \in \Omega$, $w \in \mathcal{W}$, and $v \in \mathcal{V}$.
\end{theorem}
\begin{proof}
Since the polynomial in~\eqref{eq:sos-condition} is SOS, it is globally nonnegative. Then, $g_b \ge s_b\, b + s_\Phi\, \Phi + s_w\bigl(1 - w^\top Q_w w\bigr) + s_v\bigl(1 - v^\top R_v v\bigr) \ge 0,$
because, for $x \in \mathcal{B}$, $e \in \Omega$, $w \in \mathcal{W}$, and $v \in \mathcal{V}$, each term is the product of a nonnegative SOS multiplier and a nonnegative factor.
\end{proof}

\subsection{Error-Safe Set Invariance Under Full Connection}

We next establish that, when all agents are in connected mode, the error-safe set $\Omega$ is forward invariant. Let $\mathcal{L}_{\mathrm{fc}}$ denote the Laplacian of the fully connected topology, in which every agent is in connected mode. Specializing the stacked error dynamics~\eqref{eq:dynamics-error} to this topology gives
\begin{equation}\label{eq:error-fc}
    \dot e(t) = A_e(\mathcal{L}_{\mathrm{fc}}) e(t) + \mathcal{I} w(t) - J(\mathcal{L}_{\mathrm{fc}}) v(t),
\end{equation}
where $A_e(\mathcal{L}_{\mathrm{fc}}) = F(\mathcal{L}_{\mathrm{fc}}) - \eta(\mathcal{L}_{\mathrm{fc}} \otimes I_n)$.

\begin{proposition}[Fully connected recovery]
\label{prop:recovery}
Let $S_e(\mathcal{L}_{\mathrm{fc}}) \triangleq A_e(\mathcal{L}_{\mathrm{fc}})^\top \bar P + \bar P A_e(\mathcal{L}_{\mathrm{fc}})$. If there exist a weighting matrix $\bar P = I_N \otimes P_e \succ 0$ and scalars $\alpha > 0$, $\beta_w, \beta_v \ge 0$ such that
\begin{subequations}\label{eq:recovery-lmi}
\begin{align}
    M(\bar P, \alpha, \beta_w, \beta_v) \triangleq
    &\scalebox{0.68}{$
    \begin{bmatrix} S_e(\mathcal{L}_{\mathrm{fc}}) + \alpha \bar P & \bar P \mathcal{I} & -\bar P J(\mathcal{L}_{\mathrm{fc}}) \\ \mathcal{I}^\top \bar P & -\beta_w Q_w & 0 \\ -J(\mathcal{L}_{\mathrm{fc}})^\top \bar P & 0 & -\beta_v R_v \end{bmatrix}$} \preceq 0, \label{eq:recovery-lmi-matrix} \\[4pt]
    &\alpha \gamma_{\mathrm{off}} > \beta_w + \beta_v, \label{eq:recovery-lmi-scalar}
\end{align}
\end{subequations}
then, along the fully connected error dynamics~\eqref{eq:error-fc},
\begin{equation}\label{eq:recovery}
    \dot V_e(e; \mathcal{L}_{\mathrm{fc}}) = 2 e^\top \bar P\bigl(A_e(\mathcal{L}_{\mathrm{fc}}) e + \mathcal{I} w - J(\mathcal{L}_{\mathrm{fc}}) v\bigr) < 0
\end{equation}
for all $e$ with $V_e(e) \ge \gamma_{\mathrm{off}}$ and all admissible $w \in \mathcal{W}$ and $v \in \mathcal{V}$.
\end{proposition}

\begin{proof}
Pre- and post-multiplying~\eqref{eq:recovery-lmi-matrix} by $[\,e^\top\ w^\top\ v^\top\,]$ and its transpose gives
\begin{equation}\label{eq:Vdot-bound}
    \dot V_e(e; \mathcal{L}_{\mathrm{fc}}) \le -\alpha\, e^\top \bar P e + \beta_w w^\top Q_w w + \beta_v v^\top R_v v.
\end{equation}
With $w^\top Q_w w \le 1$ and $v^\top R_v v \le 1$; since $e^\top \bar P e = V_e$, this gives $\dot V_e \le -\alpha V_e + \beta_w + \beta_v$. Whenever $V_e \ge \gamma_{\mathrm{off}}$, we have $-\alpha V_e \le -\alpha\gamma_{\mathrm{off}}$, so $\dot V_e \le -\alpha\gamma_{\mathrm{off}} + \beta_w + \beta_v < 0,$ where the strict inequality is the scalar condition~\eqref{eq:recovery-lmi-scalar}. This establishes the strict decrease~\eqref{eq:recovery}.
\end{proof}

% \begin{remark}
% Condition~\eqref{eq:recovery} requires strict decrease of the error only outside the re-entry set $\{e : V_e(e) \le \gamma_{\mathrm{off}}\}$. Decrease cannot hold on all of $\mathbb{R}^{Nn}$: near $e = 0$ the disturbance terms dominate the quadratic term and $\dot V_e$ may be positive. The annulus $V_e \ge \gamma_{\mathrm{off}}$ is bounded away from the origin, removing this obstruction.
% \end{remark}

\begin{lemma}
\label{lem:E-invariant}
Suppose every agent is in connected mode on the time interval $[t_1, t_2]$ and the recovery condition~\eqref{eq:recovery} holds. If $e(t_1) \in \Omega$, then $e(t) \in \Omega$ for all $t \in [t_1, t_2]$.
\end{lemma}
\begin{proof}
The set $\Omega = \{e : V_e(e) \le 1\}$ is a sublevel set of $V_e$, so it is forward invariant provided $V_e$ does not increase through its boundary $\partial\Omega = \{e : V_e(e) = 1\}$. On $\partial\Omega$, $V_e(e(t)) = 1 \ge \gamma_{\mathrm{off}}$, so~\eqref{eq:recovery} gives $\dot V_e(e(t); \mathcal{L}_{\mathrm{fc}}) < 0$. Hence a trajectory reaching $\partial\Omega$ has strictly decreasing $V_e$ and cannot cross from $V_e \le 1$ to $V_e > 1$. With $e(t_1) \in \Omega$, it follows that $V_e(e(t)) \le 1$ for all $t \in [t_1, t_2]$.
\end{proof}

\subsection{Safety Guarantees}

We now combine the preceding results to guarantee safety under the switching protocol. As long as the estimation error remains in $\Omega$, the state stays safe.

\begin{proposition}
\label{prop:safety-core}
Suppose $x(t_1) \in \mathcal{B}$, $\mathcal{B} \subseteq \mathcal{C}$, condition~\eqref{eq:sos-condition} holds, and $e(t) \in \Omega$ for all $t \in [t_1, t_2]$. Then $x(t) \in \mathcal{C}$ for all $t \in [t_1, t_2]$.
\end{proposition}
\begin{proof}
Since $e(t) \in \Omega$ on $[t_1, t_2]$, Theorem~\ref{thm:sos-cbf} gives the robust CBF condition~\eqref{eq:cbf-closedloop}. So, along the closed-loop dynamics~\eqref{eq:closed_loop_dynamics}, $\dot b(x(t)) \ge -\alpha_b(b(x(t)))$ whenever $x(t) \in \mathcal{B}$. With $x(t_1) \in \mathcal{B}$, we have $b(x(t_1)) \ge 0$; since $\alpha_b(0) = 0$, $\dot b \ge 0$ on the boundary $\{x:b(x) = 0\}$, so $b(x(t))$ cannot become negative. Hence $x(t) \in \mathcal{B}$, and since $\mathcal{B} \subseteq \mathcal{C}$, $x(t) \in \mathcal{C}$ for all $t \in [t_1, t_2]$.
\end{proof}

By Proposition~\ref{prop:safety-core}, safety follows once $e(t) \in \Omega$ is maintained for all $t$. Since the network is partially or fully connected at every instant, these two cases cover all $t \ge 0$; ensuring $e(t) \in \Omega$ in each therefore guarantees safety throughout.

\begin{theorem}
\label{thm:switching}
Let $b$ and $h$ be polynomials satisfying $h(x) - b(x) \in \Sigma[x]$, and suppose~\eqref{eq:sos-condition} holds. Suppose the system is initialized with $x(0) \in \mathcal{B}$ and $e(0) \in \Omega$, and that the time interval $[0, \infty)$ is partitioned into intervals, each of which is either partially connected or fully connected:
\begin{itemize}
\item Partially connected interval: for every $t$ in the interval, there exists an agent $i\in\mathcal N$ such that $i\notin\Psi(t)$.
\item Fully connected interval: every agent is in connected mode throughout the interval, and the recovery condition~\eqref{eq:recovery} holds.
\end{itemize}
Then:
\begin{enumerate}
    \item[(i)] the estimation error satisfies $e(t) \in \Omega$ for all $t \ge 0$;
    \item[(ii)] the set $\mathcal{B}$ is forward invariant, and since $\mathcal{B} \subseteq \mathcal{C}$, the state satisfies $x(t) \in \mathcal{C}$ for all $t \ge 0$.
\end{enumerate}
\end{theorem}
\begin{proof}
\emph{(i).} We show $e(t) \in \Omega$ for all $t \ge 0$ by induction over the intervals of the partition. Let the intervals be $[t_k, t_{k+1})$ for $k = 0, 1, \dots$ with $t_0 = 0$, and suppose $e(t_k) \in \Omega$; the base case $e(t_0) = e(0) \in \Omega$ holds by assumption.
If $[t_k, t_{k+1})$ is partially connected, then for every $t$ in the interval there exists an agent $i \in \mathcal{N}$ with $i\notin\Psi(t)$. By~\eqref{eq:trigger-rule}, such an agent must satisfy $U_i(t)\le1$; otherwise it would connect. Since $V_e(e(t))\le U_i(t)$, we have $V_e(e(t))\le1$, and hence $e(t)\in\Omega$. If $[t_k, t_{k+1})$ is fully connected, then every agent is in connected mode and the recovery condition~\eqref{eq:recovery} holds; since $e(t_k) \in \Omega$, Lemma~\ref{lem:E-invariant} gives $e(t) \in \Omega$ throughout the interval.
In either case $e(t) \in \Omega$ on $[t_k, t_{k+1})$, and by continuity $e(t_{k+1}) \in \Omega$, which advances the induction. Hence $e(t) \in \Omega$ for all $t \ge 0$.
\emph{(ii).} Since $h(x) - b(x) \in \Sigma[x]$, it is globally nonnegative, so $h(x) \ge b(x)$ for all $x$; hence $b(x) \ge 0$ implies $h(x) \ge 0$, i.e., $\mathcal{B} \subseteq \mathcal{C}$. By part (i), $e(t) \in \Omega$ for all $t \ge 0$, so Proposition~\ref{prop:safety-core} applied on $[0, t]$ for arbitrary $t$ yields $x(t) \in \mathcal{B}$ for all $t \ge 0$; that is, $\mathcal{B}$ is forward invariant. Since $\mathcal{B} \subseteq \mathcal{C}$, it follows that $x(t) \in \mathcal{C}$ for all $t \ge 0$.
\end{proof}

%% file: Trigger.tex
\section{Triggering Protocol}
\label{sec:trigger}

This section constructs the local certificate $U_i(t)$. Each agent must decide when to connect and disconnect using only the information available to it at time $t$, without access to the global estimation error $e(t)$ or the full topology. To enable this decision, each agent maintains a computable certificate $U_i(t)$ that upper-bounds the global error certificate $V_e(e(t))$. The certificate combines a measurable residual, which captures the component of the error observable to agent $i$, with a comparison radius that bounds the unobservable component under all topologies consistent with agent $i$'s local information. We first derive a comparison bound for the error certificate under an arbitrary admissible topology. We then compute conservative comparison constants over the topology set available to each agent and use them to evolve the comparison radius. Finally, we combine this radius with agent $i$'s residual to obtain $U_i(t)$. The certificate has the same form in the connected and disconnected modes, while the topology set used to compute the comparison constants changes with the agent's available information.

\vspace{-0.5em}

\subsection{Comparison Bound and the Error Certificate}
\label{subsec:comparison-and-certificate}

To bound the growth of $V_e$ forward in time, we evolve the comparison radius
$\bar r_e$ by a scalar quantity. The certificate rate $\dot V_e(e; \mathcal{L})$
is governed by how fast $A_e(\mathcal{L})$ amplifies the error and how strongly the
disturbances $w, v$ force it. We quantify these two effects by the growth rate
$\mu(\mathcal{L})$ and the disturbance forcing $c(\mathcal{L})$,
\begin{align}
    \mu(\mathcal{L}) &\triangleq \lambda_{\max}\bigl(\bar P^{-1/2}(A_e(\mathcal{L})^\top \bar P + \bar P A_e(\mathcal{L})) \bar P^{-1/2}\bigr), \label{eq:mu_L} \\
    c(\mathcal{L}) &\triangleq \bigl\|\bar P^{1/2}\mathcal{I} Q_w^{-1/2}\bigr\|_2 + \bigl\|\bar P^{1/2} J(\mathcal{L}) R_v^{-1/2}\bigr\|_2. \label{eq:c_L}
\end{align}

\begin{lemma}[Comparison bound]
\label{lem:comparison}
Along~\eqref{eq:dynamics-error}, the error-certificate radius $r_e(t) \triangleq \sqrt{V_e(e(t))}$ satisfies
\begin{equation}\label{eq:radius-bound}
    \dot r_e(t) \le \tfrac{1}{2}\mu(\mathcal{L}(t))\, r_e(t) + c(\mathcal{L}(t)).
\end{equation}
Let $(\mu, c)$ be constants that upper-bound $\mu(\mathcal{L}(t))$ and $c(\mathcal{L}(t))$ on an interval starting at $t_\star$, and let the comparison radius $\bar r_e(t)$ solve
\begin{equation}\label{eq:rbar-ode}
    \dot{\bar r}_e(t) = \tfrac{1}{2}\mu\, \bar r_e(t) + c.
\end{equation}
If $\bar r_e(t_\star) = r_e(t_\star)$, then $\bar r_e(t)$ dominates $r_e(t)$ thereafter,
\begin{equation}\label{eq:rbar-dominates}
    r_e(t) \le \bar r_e(t), \qquad \forall t \ge t_\star,
\end{equation}
where $\bar r_e(t)$ is given in closed form by
\begin{equation}
    \bar r_e(t) =
    \scalebox{0.9}{$
    \begin{cases}
        e^{\frac{\mu}{2}(t - t_\star)}\bar r_e(t_\star) + \dfrac{2c}{\mu}\bigl(e^{\frac{\mu}{2}(t - t_\star)} - 1\bigr), & \mu \neq 0, \\[6pt]
        \bar r_e(t_\star) + c\,(t - t_\star), & \mu = 0.
    \end{cases}$}
    \label{eq:rbar-solution}
\end{equation}
\end{lemma}

\begin{proof}
Differentiating $V_e = e^\top \bar P e$ along~\eqref{eq:dynamics-error},
\begin{equation}\label{eq:Vdot-raw}
    \dot V_e(e;\mathcal{L}) = e^\top\bigl(A_e^\top \bar P + \bar P A_e\bigr)e
    + 2e^\top \bar P \mathcal{I} w - 2 e^\top \bar P J v.
\end{equation}
By~\eqref{eq:mu_L}, the quadratic term satisfies $e^\top(A_e^\top \bar P + \bar P A_e)e
\le \mu(\mathcal{L})\, e^\top \bar P e = \mu(\mathcal{L}) V_e$. For the process-disturbance
term, writing $\bar P = \bar P^{1/2}\bar P^{1/2}$ and inserting $Q_w^{-1/2}Q_w^{1/2}=I$, and by
the Cauchy--Schwarz,
\begin{equation}\label{eq:w-cs}
\begin{aligned}
    2 e^\top \bar P \mathcal{I} w
    &= 2\bigl(\bar P^{1/2} e\bigr)^\top \bigl(\bar P^{1/2}\mathcal{I} Q_w^{-1/2}\bigr)\bigl(Q_w^{1/2} w\bigr) \\
    &\le 2\sqrt{V_e}\,\bigl\|\bar P^{1/2}\mathcal{I} Q_w^{-1/2}\bigr\|_2,
\end{aligned}
\end{equation}
using $\|\bar P^{1/2}e\|_2 = \sqrt{V_e}$ and $\|Q_w^{1/2} w\|_2 = \sqrt{w^\top Q_w w} \le 1$.
The term in $v$ is bounded identically, and their sum is $2 c(\mathcal{L})\sqrt{V_e}$
by~\eqref{eq:c_L}, so
\begin{equation}\label{eq:Vdot-mu-c}
    \dot V_e(e;\mathcal{L}) \le \mu(\mathcal{L}) V_e + 2 c(\mathcal{L}) \sqrt{V_e}.
\end{equation}
For $e \neq 0$, dividing~\eqref{eq:Vdot-mu-c} by $2\sqrt{V_e}$ and using
$\dot r_e = \dot V_e/(2\sqrt{V_e})$ gives~\eqref{eq:radius-bound}. Since $(\mu, c)$
upper-bound $(\mu(\mathcal{L}(t)), c(\mathcal{L}(t)))$, the right-hand side
of~\eqref{eq:radius-bound} is at most $\tfrac{1}{2}\mu\, r_e(t) + c$. With
$\bar r_e(t_\star) = r_e(t_\star)$, the comparison principle applied
to~\eqref{eq:rbar-ode} gives~\eqref{eq:rbar-dominates}, and
integrating~\eqref{eq:rbar-ode} gives~\eqref{eq:rbar-solution}.
\end{proof}

We now construct $U_i(t)$. Agent $i$ forms the measurable residual $\delta_i(t) \triangleq y_i(t) - C_i \hat x_i(t) = C_i e_i(t) + v_i(t)$. With the block-selection matrix $\Pi_i$ satisfying $e_i = \Pi_i e$, the lifted matrix $\mathcal{C}_i \triangleq C_i \Pi_i$, and the noise-free residual $\zeta_i \triangleq \delta_i - v_i = \mathcal{C}_i e$, the Moore--Penrose pseudoinverse $\mathcal{C}_i^\dagger$ and $\Lambda_i \triangleq I - \mathcal{C}_i^\dagger \mathcal{C}_i$ split the error as $e(t) = \mathcal{C}_i^\dagger \zeta_i(t) + \Lambda_i e(t).$ Then,
\begin{equation}\label{eq:Ve-expansion}
\begin{aligned}
    V_e(e) = e^\top \bar P e = {}& \underbrace{\zeta_i^\top (\mathcal{C}_i^\dagger)^\top \bar P\, \mathcal{C}_i^\dagger \zeta_i}_{\text{(I) measurable}}
    + \underbrace{2\zeta_i^\top (\mathcal{C}_i^\dagger)^\top \bar P \Lambda_i e}_{\text{(II) cross}} \\
    & + \underbrace{(\Lambda_i e)^\top \bar P (\Lambda_i e)}_{\text{(III) hidden}}.
\end{aligned}
\end{equation}
\begin{align}
    Y_i &\triangleq 2(\mathcal{C}_i^\dagger)^\top \bar P\, \mathcal{C}_i^\dagger \succeq 0, \label{eq:Yi} \\
    \rho_i &\triangleq \lambda_{\max}\bigl(\bar P^{-1/2}\Lambda_i^\top \bar P \Lambda_i \bar P^{-1/2}\bigr). \label{eq:rhoi}
\end{align}
Together with the noise bound $v_i(t)^\top R_i v_i(t) \le 1$, $R_i \succ 0$, these yield the certificate.

\begin{proposition}
\label{prop:Ui}
Let $\bar r_e(t)$ defined in \eqref{eq:rbar-solution} be a comparison radius with $r_e(t) \le \bar r_e(t)$, and define
\begin{equation}\label{eq:Ui}
\begin{aligned}
    U_i(t) \triangleq {}& 2\delta_i(t)^\top Y_i \delta_i(t) + 2\lambda_{\max}\bigl(R_i^{-1/2} Y_i R_i^{-1/2}\bigr) \\
    & + 2\rho_i\, \bar r_e(t)^2,
\end{aligned}
\end{equation}
then $V_e(e(t)) \le U_i(t)$.
\end{proposition}
\begin{proof}
We bound the three terms of~\eqref{eq:Ve-expansion} separately.

\emph{Term (I).} By the definition~\eqref{eq:Yi} of $Y_i$,
\begin{equation}\label{eq:termI}
    \zeta_i^\top (\mathcal{C}_i^\dagger)^\top \bar P\, \mathcal{C}_i^\dagger \zeta_i = \tfrac{1}{2}\zeta_i^\top Y_i \zeta_i.
\end{equation}

\emph{Term (II).} By the Cauchy--Schwarz inequality and then Young's inequality $2ab \le a^2 + b^2$,
\begin{equation}\label{eq:cross}
\begin{aligned}
    2\zeta_i^\top (\mathcal{C}_i^\dagger)^\top \bar P \Lambda_i e
    &\le 2\bigl\|\bar P^{1/2}\mathcal{C}_i^\dagger \zeta_i\bigr\|_2 \bigl\|\bar P^{1/2}\Lambda_i e\bigr\|_2 \\
    &\le \bigl\|\bar P^{1/2}\mathcal{C}_i^\dagger \zeta_i\bigr\|_2^2 + \bigl\|\bar P^{1/2}\Lambda_i e\bigr\|_2^2.
\end{aligned}
\end{equation}

\emph{Term (III).} Writing $\xi \triangleq \bar P^{1/2} e$ and using the definition~\eqref{eq:rhoi} of $\rho_i$,
\begin{equation}\label{eq:hidden}
    (\Lambda_i e)^\top \bar P(\Lambda_i e) = \xi^\top \bar P^{-1/2}\Lambda_i^\top \bar P \Lambda_i \bar P^{-1/2}\xi \le \rho_i V_e \le \rho_i \bar r_e^2,
\end{equation}
where the last step uses $r_e \le \bar r_e$.

The squared norms in~\eqref{eq:cross} equal $\zeta_i^\top (\mathcal{C}_i^\dagger)^\top \bar P\, \mathcal{C}_i^\dagger \zeta_i$ and $(\Lambda_i e)^\top \bar P(\Lambda_i e)$, so substituting~\eqref{eq:termI}--\eqref{eq:hidden} into~\eqref{eq:Ve-expansion} gives
\begin{equation}\label{eq:Ve-combined}
\begin{aligned}
    V_e(e) &\le 2\zeta_i^\top (\mathcal{C}_i^\dagger)^\top \bar P\, \mathcal{C}_i^\dagger \zeta_i + 2(\Lambda_i e)^\top \bar P(\Lambda_i e) \\
    &= \zeta_i^\top Y_i \zeta_i + 2(\Lambda_i e)^\top \bar P(\Lambda_i e).
\end{aligned}
\end{equation}
The remaining noise in $\zeta_i = \delta_i - v_i$ is removed using $Y_i \succeq 0$. Then,
\begin{equation}\label{eq:noise-young}
\begin{aligned}
    \zeta_i^\top Y_i \zeta_i
    &= \delta_i^\top Y_i \delta_i + \underbrace{\bigl(- 2\delta_i^\top Y_i v_i\bigr)}_{\le\, \delta_i^\top Y_i \delta_i + v_i^\top Y_i v_i} + v_i^\top Y_i v_i \\
    &\le 2\delta_i^\top Y_i \delta_i + 2 v_i^\top Y_i v_i,
\end{aligned}
\end{equation}
and the noise bound $v_i^\top R_i v_i \le 1$ gives
\begin{equation}\label{eq:noise-bound}
    v_i^\top Y_i v_i \le \lambda_{\max}\bigl(R_i^{-1/2} Y_i R_i^{-1/2}\bigr).
\end{equation}
With ~\eqref{eq:noise-young} and~\eqref{eq:noise-bound} in~\eqref{eq:Ve-combined}, $V_e(e(t)) \le U_i(t)$.
\end{proof}

The certificate is computable from the information available to agent $i$ at time $t$: the residual $\delta_i(t) = y_i(t) - C_i \hat x_i(t)$ uses only agent $i$'s measurement and estimate; $Y_i, \rho_i, R_i$ are known design quantities; and $\bar r_e(t)$ follows from~\eqref{eq:rbar-solution} given the comparison constants $(\mu, c)$. The two triggers differ only in the constants $(\mu, c)$ used to evolve $\bar r_e(t)$.

\subsection{Mode-Dependent Triggering Certificate}
\label{subsec:mode-dependent-trigger}

Lemma~\ref{lem:comparison} provides a comparison bound for any admissible
topology. Thus the same local certificate is used in both modes; only the
topology set over which the comparison constants are computed changes.

Let $\mathfrak L(\bar{\mathcal G})$ denote the set of Laplacians induced by
subgraphs of the underlying communication graph $\bar{\mathcal G}$. Since
$\bar{\mathcal G}$ has finitely many subgraphs, $\mathfrak L(\bar{\mathcal G})$
is a finite set. For agent $i$, let
$\mathcal H_i^\star(t)\subseteq\mathfrak L(\bar{\mathcal G})$ collect the
admissible Laplacians consistent with the topology information available to
agent $i$ at time $t$, and define
\begin{equation}
\label{eq:mode-dependent-constants}
\mu_i^\star(t) \triangleq \sup_{\mathcal L\in\mathcal H_i^\star(t)}\mu(\mathcal L),
\qquad
c_i^\star(t) \triangleq \sup_{\mathcal L\in\mathcal H_i^\star(t)}c(\mathcal L),
\end{equation}
where $\mu(\mathcal L)$, $c(\mathcal L)$ are defined
in~\eqref{eq:mu_L}--\eqref{eq:c_L}. Since each agent is either connected or disconnected, there are at most $2^N$ topology candidates. Thus, each supremum is computable over the candidates using agent $i$'s information.
Because the true
topology satisfies $\mathcal L(t)\in\mathcal H_i^\star(t)$, evolving
$\bar r_e(t)$ by~\eqref{eq:rbar-solution} with $(\mu_i^\star(t),c_i^\star(t))$
guarantees $r_e(t)\le\bar r_e(t)$ for every topology consistent with agent $i$'s
information. The local certificate is therefore~\eqref{eq:Ui}, and by
Proposition~\ref{prop:Ui}, $V_e(e(t))\le U_i(t)$.

(i) When agent $i$ is disconnected, it has no active neighbors, so $\ell_{ii}(t)=0$,
but it does not know the rest of the network. Hence the admissible topology set
\begin{equation}
\label{eq:Hdis}
\mathcal H_i^\star(t)=\mathcal H_i^{\mathrm{dis}}\triangleq
\bigl\{\,\mathcal L\in\mathfrak L(\bar{\mathcal G}):\ \ell_{ii}=0\,\bigr\}
\end{equation}
is time-independent, so $(\mu_{\mathrm{dis}},c_{\mathrm{dis}})$ where $\mu_{\mathrm{dis}} \triangleq \sup_{\mathcal{L} \in \mathcal{H}_i^{\mathrm{dis}}} \mu\bigl(\mathcal{L}\bigr)$ and $c_{\mathrm{dis}}   \triangleq \sup_{\mathcal{L} \in \mathcal{H}_i^{\mathrm{dis}}} c\bigl(\mathcal{L}\bigr)$ are constants
computed offline. Since $V_e(e(t))\le U_i(t)$, the condition $U_i(t)\le 1$
implies $e(t)\in\Omega$. A disconnected agent therefore connects when $U_i(t)$
reaches the safety threshold $1$, restoring communication before the error
certificate can leave $\Omega$.

(ii) When agent $i$ is connected, it knows the topology among the agents reachable
from $i$ over active links; let $\mathcal L_i^{\mathrm{known}}(t)$ be this known
partial Laplacian. The admissible topology set is then
\begin{align*}
&\mathcal H_i^\star(t)=\mathcal H_i^{\mathrm{con}}(t)\triangleq
\Bigl\{\,\mathcal L=\mathcal L_i^{\mathrm{known}}(t)+\mathcal L_i^{\mathrm{unknown}}: \\
&\mathcal L=\mathcal L^\top, \mathcal L\mathbf 1=\mathbf 0,\
\ell_{jk}\le 0\ (j\neq k),\
\ell_{jk}=0\ \ \forall (j,k)\notin\bar{\mathcal E}\,\Bigr\},
\end{align*}

and $(\mu_i^{\mathrm{con}}(t),c_i^{\mathrm{con}}(t))$ where $\mu_i^{\mathrm{con}}(t) \triangleq \sup_{\mathcal{L} \in \mathcal{H}_i^{\mathrm{con}}(t)} \mu\bigl(\mathcal{L}\bigr)$ and $c_i^{\mathrm{con}}(t)   \triangleq \sup_{\mathcal{L} \in \mathcal{H}_i^{\mathrm{con}}(t)} c\bigl(\mathcal{L}\bigr)$ are recomputed online as the
reachable set changes. The online computation can be reduced by following the offline lookup-table approach of~\cite{10360262}: we precompute $\mu(L)$ and $c(L)$ for all at most $2^N$ topology candidates and store the results in a table for online lookup. The certificate still has the form~\eqref{eq:Ui}; only the
evolution of $\bar r_e(t)$ changes. Once $U_i(t)\le\gamma_{\mathrm{off}}$, agent
$i$ sets $\sigma_i=1$; since $\gamma_{\mathrm{off}}<1$, this implies
$e(t)\in\Omega_{\mathrm{off}}$ of~\eqref{eq:inner-set}, a margin strictly inside the error-safe set.

The certificate $U_i(t)$ provides the local quantity needed
to implement the protocol in Definition~\ref{def:thresholds}. By Proposition~\ref{prop:Ui},
$U_i(t) \le \gamma_{\mathrm{off}}$ implies
$e(t)\in\Omega_{\mathrm{off}}\subset\Omega$, so the inner threshold
creates a certified recovery margin. The synchronized release rule preserves
this margin by allowing all agents in $R_i(t)$ to disconnect only at an instant
when $U_j(t)\le\gamma_{\mathrm{off}}$ holds for every $j\in R_i(t)$.

\begin{lemma}[Non-Zeno behavior]
\label{lem:no-zeno}
Let $s^\star=\sqrt{\gamma_{\mathrm{off}}}$. The finite comparison constants used to construct $U_i(t)$, including the growth rate $\mu$ and the forcing term $c$, imply that $U_i(t)$ has a bounded rate of change on each connected or disconnected interval. Since $\gamma_{\mathrm{off}}\in(0,1)$, the threshold separation $1-\gamma_{\mathrm{off}}>0$. Therefore, there exist positive crossing times $T_{\mathrm{off}}>0$ and $T_{\mathrm{on}}(s^\star)>0$ such that, for every agent $i$,
\begin{itemize}
\item[(i)] after a disconnection at $t_{\mathrm d}$ with $U_i(t_{\mathrm d})\le\gamma_{\mathrm{off}}$,
$U_i(t)<1,~ t\in[t_{\mathrm d},\,t_{\mathrm d}+T_{\mathrm{off}});$
\item[(ii)] after a connection at $t_{\mathrm c}$ with $U_i(t_{\mathrm c})\ge1$,
$U_i(t)>\gamma_{\mathrm{off}},~
t\in[t_{\mathrm c},\,t_{\mathrm c}+T_{\mathrm{on}}(s^\star)).$
\end{itemize}
Let $\{t_k^i\}_{k\in\mathbb Z_{\ge0}}$ denote the switching instants of agent $i$. Then,
$t_{k+1}^i-t_k^i\ge T_{\min}
\triangleq \min\{T_{\mathrm{off}},\,T_{\mathrm{on}}(s^\star)\}>0 .$
Hence no agent switches infinitely often in finite time.
\end{lemma}
 
\begin{remark}
This safety requirement differs from~\cite{10360262}, where temporary error growth outside the target region can be tolerated because the objective is asymptotic convergence. For safety certification, any transient violation of $\Omega$ may invalidate the CBF certificate and therefore requires a fundamentally different, safety-driven triggering design. The proposed protocol reconnects agents before their certificates reach the safety boundary and releases connected components only after recovering to a tighter safety region, thereby maintaining $e(t)\in\Omega$ and preserving $x(t)\in\mathcal C$ for all time.
\end{remark}

%% file: Simulation.tex
\section{Simulation}
\label{sec:simulation}
\begin{figure}[ht]\centering
\includegraphics[width=1\columnwidth]{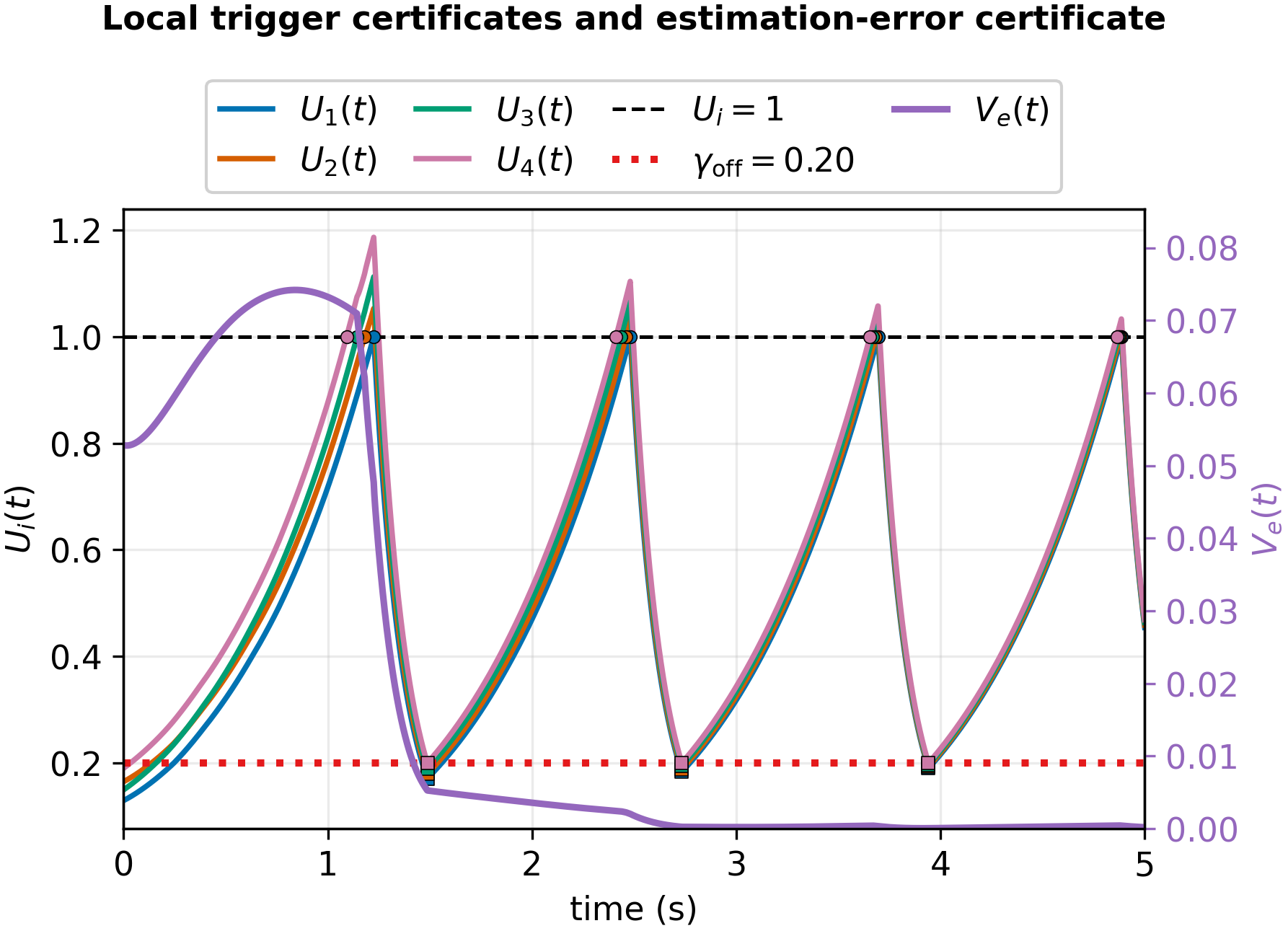}
\caption{Local triggering certificates $U_i(t)$ and the global error certificate
$V_e(t)$. Each agent connects when $U_i(t)=1$ and is ready to disconnect
whenever $U_i(t)\le\gamma_{\mathrm{off}}=0.20$. The plotted $V_e(t)$ remains
below $1$, verifying that the global estimation error stays inside the
error-safe set defined in~\eqref{eq:error_safe_set}.}
\label{fig:ui_ve}\end{figure}

\begin{figure}[ht]\centering
\includegraphics[width=0.9\columnwidth]{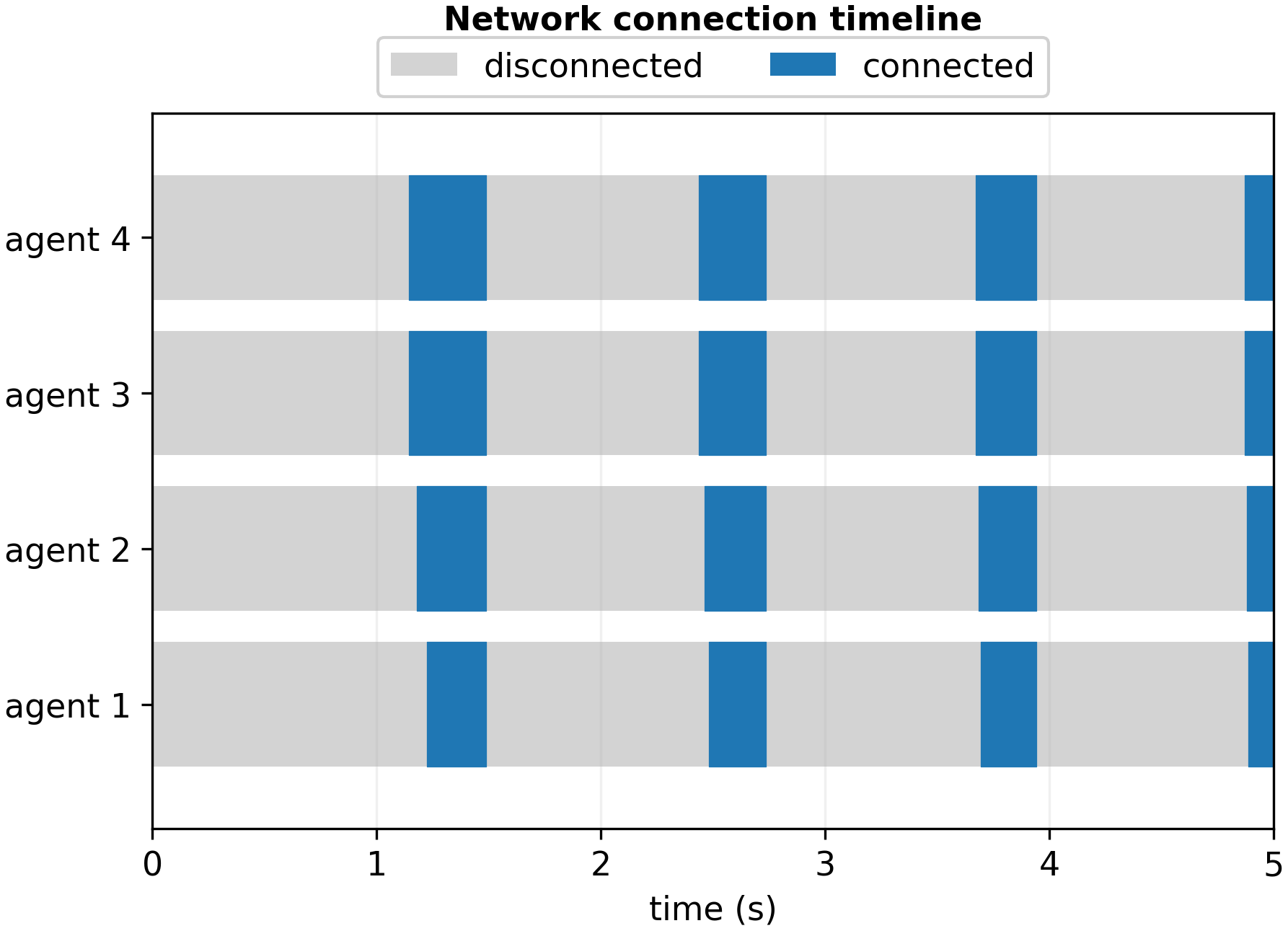}
\caption{Connected and disconnected intervals of the four agents over the
$5$\,s run. The agents are connected for only 19.6\%, 20.6\%, 21.5\%, and 21.5\% of the time. Each agent remains in disconnected mode for roughly 80\% of the time.}
\label{fig:timeline}\end{figure}

We use the quadruple-tank setup of~\cite{johansson2000}, operated about a
reference level of $3$\,m, and let $x_i$ denote the deviation of tank $i$ from
this reference. Our goal is to verify that, under the triggering protocol, the
four agents keep all tank levels within a prescribed safe band while
communicating only intermittently. Safety requires each deviation to satisfy
$|x_i|\le2$\,m, i.e.\ the safe set is $\mathcal C=\{x:\|x\|_\infty\le2\}$; we
certify forward invariance of the inscribed ellipsoid
$\mathcal B=\{x \in \mathbb{R}^4:x^\top H_bx\le1\}$ with $H_b=\mathrm{diag}(1/4)$, so
$\mathcal B\subset\mathcal C$, and the initial deviation
$x(0)=[0.6,1.0,-0.5,0.3]$\,m lies in $\mathcal B$. Each agent senses and actuates its own tank, so $C=I_4$ and $B$ is diagonal;
linearizing about the reference level
gives $\dot x=Ax+Bu$, $y_i=C_ix+v_i$ with
\begin{equation}
\begin{aligned}
A&=\begin{bmatrix}
-0.0624 & 0.0100 & 0.0100 & 0.0100\\
 0.0100 & -0.0528 & 0.0100 & 0.0100\\
 0.0100 & 0.0100 & -0.0624 & 0.0100\\
 0.0100 & 0.0100 & 0.0100 & -0.0528
\end{bmatrix},\\[2pt]
B&=\mathrm{diag}(0.1189,0.1047,0.1189,0.1047).
\end{aligned}
\label{eq:Amat}
\end{equation}
The open loop is stable, with eigenvalues
$\mathrm{eig}(A)=\{-0.027,-0.063,-0.068,-0.072\}$. The feedback $u_i=K_i\hat x_i$ is chosen by placing the nominal closed-loop poles at $-1.5$, so that $A+BK=-1.5I$. The communication graph is complete, connected agents run a full-state observer
with $A-LC=-4I$ and consensus gain $\eta=50$. Disconnected agents use a local observer with pole at $-0.4$. The disturbance and noise satisfy
$\|w\|\le0.012$\,m/s and $\|v\|\le0.012$\,m, i.e. $Q_w=R_v=(1/0.012^2)I$.
Subject to $h(x)-b(x)\in\Sigma[x]$, $b(x(0))\ge 0$, the SOS condition~\eqref{eq:sos-condition}, and the recovery LMI~\eqref{eq:recovery-lmi}, with $\bar P$ structured as in Definition~\ref{def:structured_error_weighting_matrix}, a feasible solution is $\bar P=I_N\otimes P_e$, where 
\begin{equation}
P_e=\begin{bmatrix}
\phantom{-}1.009 & -0.006 & -0.006 & -0.006\\
-0.006 & \phantom{-}0.991 & -0.006 & -0.006\\
-0.006 & -0.006 & \phantom{-}1.009 & -0.006\\
-0.006 & -0.006 & -0.006 & \phantom{-}0.991
\end{bmatrix}.
\label{eq:Pe}
\end{equation}
We take the release threshold
$\gamma_{\mathrm{off}}=0.20$. Running the protocol from $x(0)$, the state stays in $\mathcal B$ with
$x^\top H_bx\le0.425$, hence in $\mathcal C$, so every deviation remains within
$\pm2$\,m, while the network is disconnected for roughly $80\%$ of the run
(Figs.~\ref{fig:ui_ve} and~\ref{fig:timeline}). This confirms that the triggering
protocol preserves both estimation-error and state safety while communicating
only intermittently.

%% file: Conclusion.tex
\section{Conclusion}
\label{sec:conclusion}
This paper developed a decentralized event-triggered topology-switching protocol for multi-agent systems under estimation uncertainty. In the intermittent setting, each agent switched between connected and disconnected modes and made decisions using locally computable certificates, while the state remained safe. The protocol reconnected agents before certificates reached the safety boundary and disconnected them only after recovery below a stricter limit. It maintained the estimation error within a certified safety bound and excluded Zeno behavior. In the quadruple-tank example, all tank deviations remained within the prescribed safety bound, with $x^\top H_b x \le 0.425$, while agents stayed disconnected for roughly $80\%$ of the run. Future work will reduce conservatism by optimizing the error weighting matrix $\bar P$ and the inner triggering threshold $\gamma_{\mathrm{off}}$.

%% file: references.bib
@ARTICLE{johansson2000,
  author={Johansson, K.H.},
  journal={IEEE Transactions on Control Systems Technology}, 
  title={The quadruple-tank process: a multivariable laboratory process with an adjustable zero}, 
  year={2000},
  volume={8},
  number={3},
  pages={456-465},
  doi={10.1109/87.845876},
  ISSN={1558-0865},
  month={May},}

@ARTICLE{10360262,
  author={Griffioen, Paul and Romagnoli, Raffaele and Krogh, Bruce H. and Sinopoli, Bruno},
  journal={IEEE Transactions on Control of Network Systems}, 
  title={Reducing Attack Opportunities Through Decentralized Event-Triggered Control}, 
  year={2024},
  volume={11},
  number={3},
  pages={1576-1586},
  doi={10.1109/TCNS.2023.3343151}}

@article{cui2023decentralized,
  title={A decentralized dynamic self-triggered control approach to consensus of multiagent systems},
  author={Cui, Hailong and Zhao, Guanglei and Liu, Shuang and Li, Zhijie},
  journal={IEEE Transactions on Systems, Man, and Cybernetics: Systems},
  volume={53},
  number={9},
  pages={5772--5783},
  year={2023},
  publisher={IEEE}
}

@book{bullo2009distributed,
  title={Distributed control of robotic networks: a mathematical approach to motion coordination algorithms},
  author={Bullo, Francesco and Cort{\'e}s, Jorge and Martinez, Sonia},
  year={2009},
  publisher={Princeton University Press}
}

@article{deng2022distributed,
  title={Distributed adaptive tracking control for high-order nonlinear multiagent systems over event-triggered communication},
  author={Deng, Chao and Wen, Changyun and Wang, Wei and Li, Xinyao and Yue, Dong},
  journal={IEEE Transactions on Automatic Control},
  volume={68},
  number={2},
  pages={1176--1183},
  year={2022},
  publisher={IEEE}
}

@article{nowzari2019event,
  title={Event-triggered communication and control of networked systems for multi-agent consensus},
  author={Nowzari, Cameron and Garcia, Eloy and Cort{\'e}s, Jorge},
  journal={Automatica},
  volume={105},
  pages={1--27},
  year={2019},
  publisher={Elsevier}
}

@article{an2023event,
  title={Event-triggered reference governors for collisions-free leader--following coordination under unreliable communication topologies},
  author={An, Liwei and Yang, Guang-Hong and Deng, Chao and Wen, Changyun},
  journal={IEEE Transactions on Automatic Control},
  volume={69},
  number={4},
  pages={2116--2130},
  year={2023},
  publisher={IEEE}
}

@article{wang2024control,
  title={Control barrier performance function-based cooperative formation with parallel dynamic event-triggering strategy},
  author={Wang, Peng and Liang, Xiaoling and Peng, Xiuhui and Lu, Yu and Ge, Shuzhi Sam},
  journal={IEEE Transactions on Systems, Man, and Cybernetics: Systems},
  volume={54},
  number={7},
  pages={4552--4564},
  year={2024},
  publisher={IEEE}
}

@article{ge2021dynamic,
  title={Dynamic event-triggered control and estimation: A survey},
  author={Ge, Xiaohua and Han, Qing-Long and Zhang, Xian-Ming and Ding, Derui},
  journal={International Journal of Automation and Computing},
  volume={18},
  number={6},
  pages={857--886},
  year={2021},
  publisher={Springer}
}

@article{rajaraman2002topology,
  title={Topology control and routing in ad hoc networks: A survey},
  author={Rajaraman, Rajmohan},
  journal={ACM SIGACT News},
  volume={33},
  number={2},
  pages={60--73},
  year={2002},
  publisher={ACM New York, NY, USA}
}

@inproceedings{sabouni2023optimal,
  title={Optimal control of connected automated vehicles with event-triggered control barrier functions: a test bed for safe optimal merging},
  author={Sabouni, Ehsan and Ahmad, HM Sabbir and Xiao, Wei and Cassandras, Christos G and Li, Wenchao},
  booktitle={2023 IEEE Conference on Control Technology and Applications (CCTA)},
  pages={321--326},
  year={2023},
  organization={IEEE}
}

@article{luo2020multi,
  title={Multi-robot collision avoidance under uncertainty with probabilistic safety barrier certificates},
  author={Luo, Wenhao and Sun, Wen and Kapoor, Ashish},
  journal={Advances in Neural Information Processing Systems},
  volume={33},
  pages={372--383},
  year={2020}
}

@article{peng2013event,
  title={Event-triggered communication and H∞ control co-design for networked control systems},
  author={Peng, Chen and Yang, Tai Cheng},
  journal={Automatica},
  volume={49},
  number={5},
  pages={1326--1332},
  year={2013},
  publisher={Elsevier}
}

@INPROCEEDINGS{8796030,
  author={Ames, Aaron D. and Coogan, Samuel and Egerstedt, Magnus and Notomista, Gennaro and Sreenath, Koushil and Tabuada, Paulo},
  booktitle={2019 18th European Control Conference (ECC)}, 
  title={Control Barrier Functions: Theory and Applications}, 
  year={2019},
  volume={},
  number={},
  pages={3420-3431},
  doi={10.23919/ECC.2019.8796030},
  ISSN={},
  month={June},}
